\documentclass{iopjournal}
\usepackage{graphicx} 

\usepackage{amsmath}
\usepackage{amssymb}
\usepackage{graphicx}
\usepackage{ulem}
\usepackage{soul}
\usepackage{xcolor}
\usepackage{siunitx}
\usepackage[colorlinks=true, allcolors=blue]{hyperref}

\usepackage{wrapfig,lipsum} 

\usepackage{tikz,lipsum,lmodern}
\usepackage[most]{tcolorbox}
\usepackage{subfig}

\newcommand{\eqstackrel}[1] {\stackrel{\substack{\ensuremath{#1} \\ \displaystyle \downarrow \\ ~}}{ = } }

\usepackage{wrapfig,lipsum} 

\usepackage{tikz,lipsum,lmodern}
\usepackage[most]{tcolorbox}
\usepackage{subfig}

\usepackage{pdflscape}

\usepackage{xcolor}
\definecolor{GreenDark}{RGB}{000,179,000}
\newcommand{\VL}[1]{\mathbf{#1}}

\usepackage{tikz}
\usepackage{tikz-3dplot}
\usepackage{pgfplots}
\pgfplotsset{compat = 1.3}
\usetikzlibrary{arrows}
\usetikzlibrary{calc}
\newcommand{\TikzInnerSep}{0.33mm}

\def\centerarc[#1](#2)(#3:#4:#5){%
	\draw [#1] ($(#2)+({#5*cos(#3)}, {#5*sin(#3)})$) arc (#3:#4:#5)%
}

\newcommand{\tdseteulerzxz}{%
	\renewcommand{\tdplotcalctransformrotmain}{%
		\tdplotsinandcos{\sinalpha}{\cosalpha}{\tdplotalpha} 
		\tdplotsinandcos{\sinbeta}{\cosbeta}{\tdplotbeta}
		\tdplotsinandcos{\singamma}{\cosgamma}{\tdplotgamma}
		\tdplotmult{\sasb}{\sinalpha}{\sinbeta}
		\tdplotmult{\sacb}{\sinalpha}{\cosbeta}
		\tdplotmult{\sacbsg}{\sacb}{\singamma}
		\tdplotmult{\sacbcg}{\sacb}{\cosgamma}
		\tdplotmult{\sasg}{\sinalpha}{\singamma}
		\tdplotmult{\sacg}{\sinalpha}{\cosgamma}
		\tdplotmult{\sbsg}{\sinbeta}{\singamma}
		\tdplotmult{\sbcg}{\sinbeta}{\cosgamma}
		\tdplotmult{\casb}{\cosalpha}{\sinbeta}
		\tdplotmult{\cacb}{\cosalpha}{\cosbeta}
		\tdplotmult{\cacbsg}{\cacb}{\singamma}
		\tdplotmult{\cacbcg}{\cacb}{\cosgamma}
		\tdplotmult{\casg}{\cosalpha}{\singamma}
		\tdplotmult{\cacg}{\cosalpha}{\cosgamma}
		\pgfmathsetmacro{\raaeul}{+\cacg - \sacbsg}
		\pgfmathsetmacro{\rabeul}{-\casg - \sacbcg}
		\pgfmathsetmacro{\raceul}{+\sasb}
		\pgfmathsetmacro{\rbaeul}{+\sacg + \cacbsg}
		\pgfmathsetmacro{\rbbeul}{-\sasg + \cacbcg}
		\pgfmathsetmacro{\rbceul}{-\casb}
		\pgfmathsetmacro{\rcaeul}{+\sbsg}
		\pgfmathsetmacro{\rcbeul}{+\sbcg}
		\pgfmathsetmacro{\rcceul}{+\cosbeta}
	}
}

\tdplotsetmaincoords{70}{150}

\pgfmathsetmacro{\zOneRot}{10}
\pgfmathsetmacro{\xRot}{15}
\pgfmathsetmacro{\zTwoRot}{30}

\pgfmathsetmacro{\xRot}{10}
\pgfmathsetmacro{\yRot}{15}
\pgfmathsetmacro{\zRot}{30}

\usepackage[
backend=biber,
style=numeric,
sorting=none
]{biblatex} 
\begin{document}

\articletype{Paper} 

\title{Macroscopic evidence of spatial modulation of conductivity in a microtextured ferromagnetic film}

\author{C.P. Quinteros$^{1,*}$\orcid{0000-0002-9361-1382}, L. Avilés-Félix$^{2,3}$\orcid{0000-0003-3285-9891}, D. Goijman$^{2,4}$\orcid{0000-0002-8263-5274}, L. Saba$^3$, D. Pérez Morelo$^{2,3}$\orcid{0000-0001-5914-5392}, L. Granja$^5$\orcid{0000-0001-6824-3306}, M. Granada$^{2,3}$\orcid{0000-0003-0757-654X}, J. Milano$^{2,3}$\orcid{0000-0002-0990-4837}}

\affil{$^1$
Instituto de Ciencias Físicas (ICIFI, UNSAM-CONICET), Martín de Irigoyen 3100, San Martín (1650), Argentina.
}%

\affil{$^2$Instituto de Nanociencia y Nanotecnología, INN (CNEA-CONICET). Centro Atómico Bariloche, S. C. de Bariloche, R4802AGP, Río Negro, Argentina.
}%

\affil{$^3$Instituto Balseiro, UNCuyo-CNEA, S. C. de Bariloche, R4802AGP, Río Negro,
Argentina.}

\affil{$^4$UNRN, Sede Andina, S. C. de Bariloche, R8400GNA, Río Negro, Argentina.}

\affil{$^5$Instituto de Nanociencia y Nanotecnología, INN (CNEA-CONICET). Centro Atómico Constituyentes, San Martín, 1650, Buenos Aires, Argentina.}

\email{cquinteros@unsam.edu.ar}



\begin{abstract}
\noindent A 75 nm-thick Fe$_{0.5}$Pt$_{0.5}$ film is a ferromagnetic metal showing striped magnetic domains in remanence at room temperature. The magnetoresistance is characterized by varying the external temperature and the in-plane magnetic field intensity, thereby affecting its magnetic structure. Qualitatively, the resistivity is well described by using the generalized Ohm's law. High-field magnetotransport properties are successfully explained considering the competition between the expected metallic behavior and the electron-magnon interaction. In the low-field condition, we size the contribution of the magnetic texture to the macroscopic magnetotransport response by introducing a new quantity ($\Delta\rho_{L,coer}$). Consistent with the microscopic modulation of the lateral conduction, low-field measurements reveal inhomogeneities attributed to the spatial distribution of ferromagnetic domains and domain walls. By carefully analyzing the macroscopic response near the coercive field, the additional contribution to the resistivity is attributed to the domain walls themselves. In fact, this term could surpass the anisotropic term at low temperatures. In summary, this study demonstrates that spatial magnetic inhomogeneities are not only macroscopically measurable but also comparable in magnitude to other regularly considered terms, mainly at low temperatures.  

\end{abstract}

\section*{Introduction}

Domain textures can be exploited as self-assembled systems with information-processing potential \cite{rieck_ferroelastic_2023,quinteros_thermal_2024}. Assimilated to complex networks capable of transmitting electrical stimuli while implementing certain types of signal modulation, they are being studied as test benches for \textit{in-materia} computing proposals \cite{milano_materia_2022}. Understood as a paradigm in which the physical material performs computation tasks, the actual interest in finding suitable candidates to process information \textit{in-materia} \cite{jaeger_toward_2023}, combined with the solid knowledge of the Fe$_{0.5}$Pt$_{0.5}$ (FePt) magnetic texture, makes it a candidate for implementing self-assemblies of conducting and switchable units. Thus, to investigate synthetic structures conceived as self-assemblies for \textit{in-materia} signal processing, the ferromagnetic texture of a FePt film is studied by analyzing its magnetoresistive response.

FePt is an archetypal ferromagnet thoroughly studied \cite{quinteros_thermal_2024,Vasquez2008,Vasquez2009,sallica_leva_magnetic_2010,alvarez_relaxation_2013,alvarez_correlation_2014,alvarez_tunable_2015}. Its magnetic response under magnetic fields at different temperatures was explained both macroscopically \cite{sallica_leva_magnetic_2010,alvarez_relaxation_2013,alvarez_correlation_2014} and microscopically \cite{alvarez_tunable_2015,quinteros_thermal_2024}. Depending on the phase, the temperature, and the film thickness, it develops a magnetic texture of striped domains with net magnetization oriented within the film plane but with out-of-plane components alternating in sign from one domain to the other \cite{alvarez_dominios_2016}. Beyond its magnetic properties, technological realizations rely on the ability to electrically detect the internal arrangement of the structure of interest using macroscopic stimuli. For this reason, magnetoresistance is a valuable tool for analyzing whether magnetically induced changes can translate into macroscopically detectable electrical signals and, consequently, determining whether those changes are suitable for electronic applications.  

Magnetoresistance (MR) in ferromagnets (FM) has a long history of research \cite{jan_galvamomagnetic_1957-1,kent_magnetoresistance_1998,viret_spin_1996,gregg_giant_1996}. Defined as $\frac{\mathrm{R(H)-R(0)}}{\mathrm{R(H)}}$, ferromagnetic metals and certain dilute alloys demonstrate a negative MR at high-fields originated from the decrease of the probability for the \textit{s-d} scattering \cite{jan_galvamomagnetic_1957-1}. Later on, due to the miniaturization of magnetic devices, attention was focused on the low-field MR of FM materials. By then, it was shown that the low-field response depends on the structure of the magnetic domains \cite{kent_magnetoresistance_1998}. However, the magnitude of the MR or the qualitative response as a result of the magnetic domain texture was unclear. Spin-scattering effects were invoked to explain the role of the domain walls and attributed to the same mechanism as that for the giant MR effect \cite{viret_spin_1996}. As in the case of multilayers, the presence of zones with different magnetization directions (domains) mediated by a spacer (domain wall) affects conductance due to a mixing of minority and majority spin channels in the wall \cite{gregg_giant_1996}.  

In this context, the renewed interest in FePt relies on exploiting the arrangement of the magnetic texture as a plausible \textit{in-materia} self-assembly for signal processing. For this reason, this work attempts to identify and quantify the changes in the magnetic domain structure by means of the magnetoresistive response to in-plane fields over a wide range of temperatures. 

\section*{Sample details and microscopic characterization}

\begin{figure}[ht!]
    \centering
    \includegraphics[width=\linewidth]{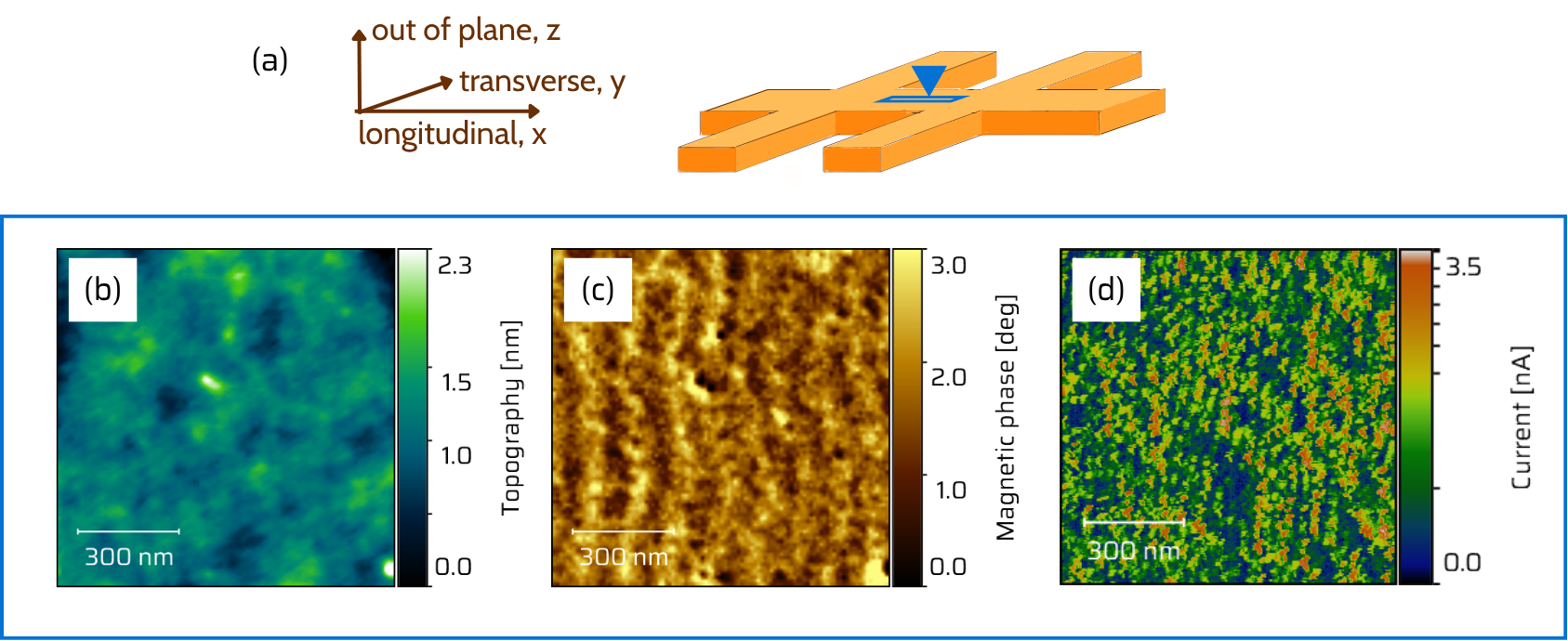}
    \caption{\textbf{Sample details.} (a) Hall bar-shaped FePt including the coordinate axes. Longitudinal and transverse directions (along and across the current injection, respectively) are contained within the film plane. The out-of-plane axis is included for completeness. The approximate location of the FePt Hall bar, where the atomic force microscopy was conducted, is sketched by a frame indicated with an arrow. (b)-(d) Atomic force microscopy images of (1$\mu$m)$^2$ at room temperature: (b) tapping-mode topography, (c) magnetic, and (d) conductive microscopy maps.}
    \label{fig:cAFM-MFM}
\end{figure}

Sputtered onto a Si substrate at room temperature (using a power density of 1.8 $\frac{\mathrm{W}}{\mathrm{cm}^2}$ in a 2.6 mTorr Ar atmosphere, resulting in a deposition rate of $\sim 0.16 \frac{\mathrm{nm}}{\mathrm{s}}$), the 75 nm-thick FePt film studied in this work exhibits a metastable crystalline phase of FCC symmetry, designated as A1 \cite{alvarez_dominios_2016}. For the magnetoresistance measurements, the film was lithographically shaped as a Hall bar (see Fig. \ref{fig:cAFM-MFM}(a)). The film geometry determines the in-plane coordinates ($\hat{x},\hat{y}$). 
For the macroscopic study, the voltage drops V$_\mathrm{L}$ and V$_\mathrm{T}$ are measured either longitudinal (parallel) or transverse (perpendicular) to the direction of current injection, i (Fig. \ref{fig:cAFM-MFM}(a)). The ratio between each voltage drop and the injected current, scaled by the corresponding area and length of the metallic path, results in the longitudinal ($\rho_L$) and transverse ($\rho_T$) resistivity terms, respectively.   

\noindent Fig. \ref{fig:cAFM-MFM} includes the microscopic characterization of the sample. Figs. \ref{fig:cAFM-MFM}(b)-(d) comprise images of atomic force microscopy at room temperature (RT) in the central part of the Hall bar (sketched as a rectangle below an arrow in Fig. \ref{fig:cAFM-MFM}(a)). Fig. \ref{fig:cAFM-MFM}(b) represents the topography of the sample. Fig. \ref{fig:cAFM-MFM}(c) displays the phase signal recorded with a magnetic tip (MESP$^{\mathrm{TM}}$ from Bruker) in a dual-pass scan (20 nm apart). Finally, Fig. \ref{fig:cAFM-MFM}(d) illustrates the current detected upon scanning a conductive tip (Etalon$^{\mathrm{TM}}$ from NanoTips) in contact mode under the application of 0.5 V. The magnetic signal is uncorrelated to the topography, as expected for the genuine magnetic texture. The MFM and cAFM maps are put together only for qualitative comparison, but cannot be mapped onto each other since the scanned windows correspond to different sample locations\footnote{This is due to limitations of the microscope to ensure the accuracy of the precise landing spot after replacing the tip used for each type of mapping.}. 
Overall, Figs. \ref{fig:cAFM-MFM}(b)-(d) demonstrate the good quality of the film, the persistence of the magnetic stripes even after the lithographic process used to shape the Hall bar, and the lateral modulation of the conductivity at the microscopic scale.  
\section*{Macroscopic magnetotransport properties}

Generalized Ohm's law \cite{harder_electrical_2016} (eq. \ref{eq:Ohm's-law}) describes the electric field due to: a current density, $\vec{J}$ (Ohm's law), an externally applied magnetic field, $\vec{H}$ (ordinary magnetoresistance) as well as the contribution of the material magnetization, $\vec{M}$ (anisotropic magnetoresistance), and the Hall effect associated to each of the two components to the magnetic induction ($\vec{H}$ produces the ordinary Hall effect while $\vec{M}$ is responsible for the anomalous Hall effect). 

\begin{equation}
    \vec{E} = \rho_{\perp} \vec{J} + \frac{\Delta \rho_0}{\vec{H}^2} (\vec{J} \cdot \vec{H}) \vec{H} + \frac{\Delta \rho}{\vec{M}^2} (\vec{J} \cdot \vec{M}) \vec{M} - \frac{\rho_H}{\mid\vec{H}\mid} \vec{J} \times \vec{H} - \frac{\rho_{AHE}}{\mid\vec{M}\mid} \vec{J} \times \vec{M} \color{black}
    \label{eq:Ohm's-law}
\end{equation}
\noindent The current injection determines the longitudinal direction, $\vec{J} = J~\hat{x}$ (see Fig. \ref{fig:cAFM-MFM}(a)), referring to the coordinate system attached to the FePt Hall bar. To deduce the expression for $\rho_L$ and $\rho_T$, 
$\rho = \frac{\hat{n} \cdot \vec{E}}{\mid\vec{J}\mid}$ is evaluated along $\hat{x}$ and $\hat{y}$, respectively. Considering the sample and field coordinate systems\footnote{The magnetization or sample coordinate system differs from the externally applied magnetic field, $\vec{H}$. This is required to successfully describe transverse resistivity $\rho_T$ when the two planes are not perfectly aligned (see Supplementary Information for details).}
, the resistivity components are expressed in equations \ref{eq:rho-x_components} and \ref{eq:rho-y_components} (see Supplementary Information for the complete equation deduction).

\begin{equation}
    \rho_L = \rho_{\perp} + \Delta\rho_0 \cdot \mathrm{\frac{H_x^2}{H^2}} + \Delta\rho \cdot \mathrm{\frac{M_x^2}{M^2}}
    \label{eq:rho-x_components}
\end{equation} 

\begin{equation}
    \rho_T = \Delta\rho_0 \cdot \mathrm{\frac{H_x \cdot H_y}{H^2}} + \Delta\rho \cdot \mathrm{\frac{M_x \cdot M_y}{M^2}} + \rho_H \mathrm{\frac{H_{z}}{H}} + \rho_{AHE} \cdot \mathrm{\frac{M_{z}}{M}}
    \label{eq:rho-y_components}
\end{equation}

\vspace{0.5cm}
\noindent The magnetotransport measurements conducted in this study include: \textbf{1.} variation of the in-plane direction of the externally applied magnetic field ($\phi_{\mathrm{H}}$) in saturation at RT, \textbf{2.} the temperature (T) variation in parallel and perpendicular magnetic field configuration either in saturation or remanence, and \textbf{3.} the variation of the magnetic field intensity (H) in the parallel and perpendicular directions for multiple T within the 80 K - 300 K range. Longitudinal ($\rho_L$) and transverse ($\rho_T$) resistivity components were measured in each condition. 

The first experiment consists of saturating the sample at RT with an in-plane magnetic field $\vec{H} = \mathrm{H}~\cdot~\cos{(\phi_{\mathrm{H}})}~\hat{x} + \mathrm{H}~\cdot~\sin{(\phi_{\mathrm{H}})}~\hat{y}$\footnote{It is worth recalling that in the deduction of the general expression, see Supplementary Information, ($\hat{x'},\hat{y'}$) coordinates were defined for the $\vec{H}$ components. This allows considering the effect in the magnetorresistance of an experimental misalignment between the planes of the sample and the rotating plate. This is not relevant for the longitudinal measurement, but it is for the transversal. For this reason, the same sample coordinates ($\hat{x},\hat{y}$) are here used to describe the field.}, being H$_\mathrm{sat} \mathrm{(RT)} = 2~$kOe $< \mathrm{H} = 8~$kOe, to observe the magnetoresistance response in the longitudinal direction as a function of the rotation angle, $\phi_{\mathrm{H}}$. The oscillating response corresponds to the Voigt-Thomson formula \cite{jan_galvamomagnetic_1957-1}. For this specific situation, since magnetization lies completely in-plane, eq. \ref{eq:rho-x_components} becomes

\begin{equation}
    \rho_L = \rho_{\perp} + [\Delta\rho + \Delta\rho_0] \cdot \cos^2(\phi_{\mathrm{H}}) 
    \label{eq:rho-L}
\end{equation} 

\begin{figure}[ht!]
    \centering
    \includegraphics[width=0.5\linewidth]{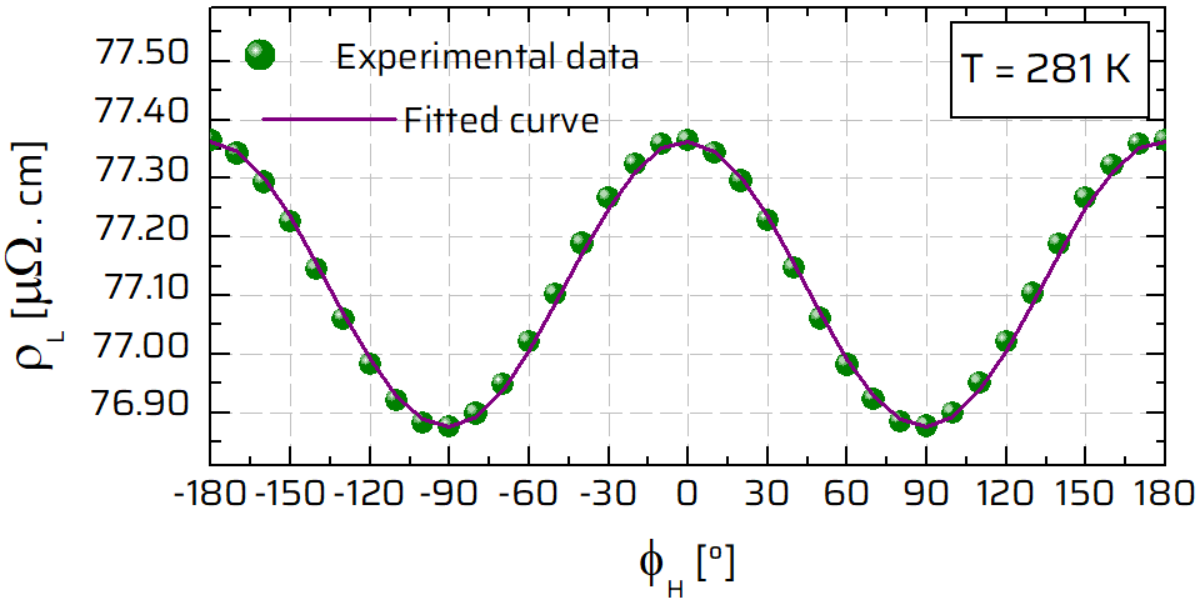}
    \caption{\textbf{Anisotropic magnetoresistance at room temperature.} 
    Longitudinal resistivity ($\rho_L$) as a function of the in-plane magnetic field direction ($\phi_{\mathrm{H}}$) measured in saturation (H = 8$~$kOe). The fitting corresponds to the Voigt-Thomson formula \cite{jan_galvamomagnetic_1957-1}, eq. \ref{eq:rho-L}.}
    \label{fig:1}
\end{figure}
\noindent By fitting the experimental data with eq. \ref{eq:rho-L}, the different resistivity terms can be deduced: $\rho_{\perp}$ = 76.8 $\mu\Omega \cdot \mathrm{cm}$ and $\Delta\rho$ + $\Delta\rho_0$ = 0.49 $\mu\Omega \cdot \mathrm{cm}$ (see Supplementary information). The obtained $\rho_{\perp}$ value is of the same order of magnitude as those reported in the literature for Fe and Pt specimens at RT ($\sim 10~\mu\Omega~\cdot$ cm) \cite{taylor_resistivity_1968-1,guan_experimental_2013} but slightly higher, as expected from the fact that the film thickness compares to the mean free path of electrons in pure metals at RT ($\sim$ tens of nm). 



The second experiment explores the resistivity ($\rho_L$) as a function of temperature (T) in different conditions: with the magnetic field applied either parallel or perpendicular to the injected current (H$//$i and H$\perp$i, respectively), measured in saturation and remanence (the latter immediately after saturating the sample).  

\begin{figure}[ht!]
    \centering
    \includegraphics[width=\linewidth]{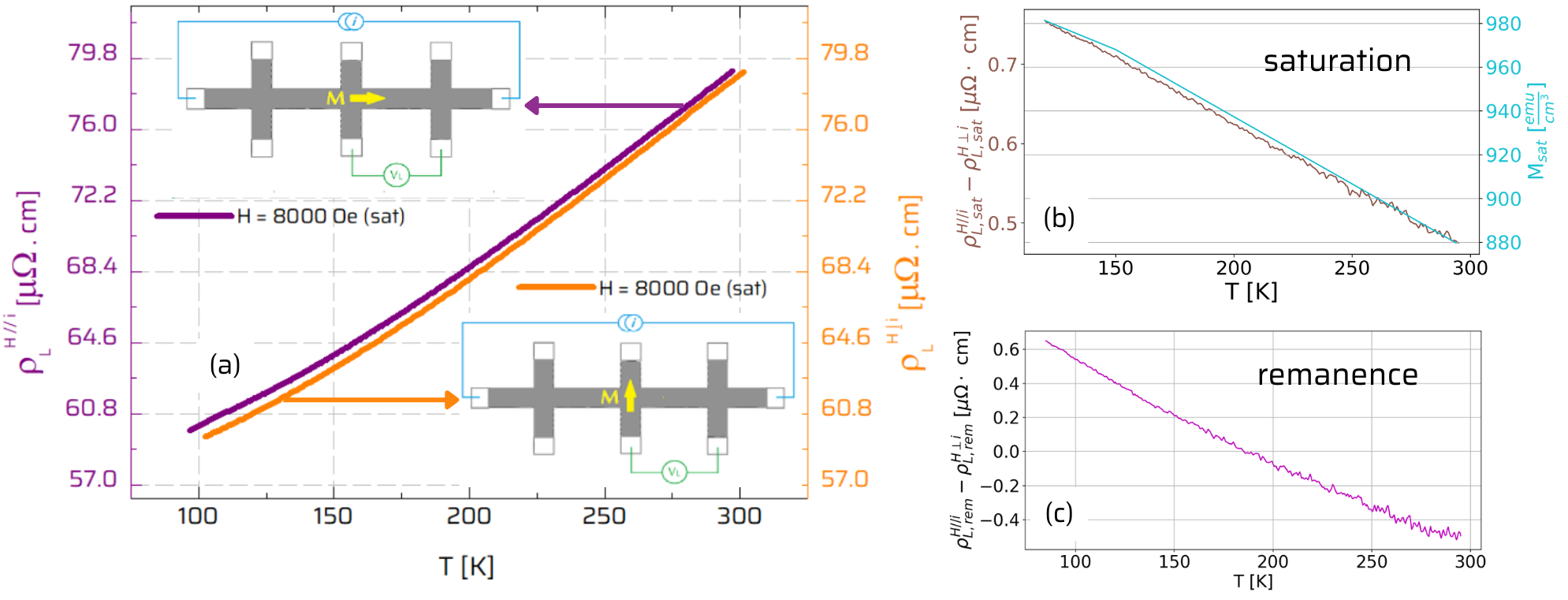}
    \caption{\textbf{Longitudinal resistivity ($\rho_L$) as a function of T.} (a) $\rho_L$ in saturation (H = 8 kOe) with H$//$i or H$\perp$i, left and right axis, respectively. (b) Difference between the two saturation resistivity measurements, $\rho_{L,sat}^{\mathrm{H}//\mathrm{i}}-\rho_{L,sat}^{\mathrm{H} \perp \mathrm{i}}$, (left axis) and magnetization (right axis) as a function of T. (c) Difference between the two remanent measurements $\rho_{L,rem}^{\mathrm{H}//\mathrm{i}}-\rho_{L,rem}^{\mathrm{H} \perp \mathrm{i}}$ as a function of T.}
    \label{fig:2}
\end{figure}

Figure \ref{fig:2}(a) presents the $\rho_L$ recorded as a function of T in saturation for H$//$i (left axis) and H$\perp$i (right axis). $\rho_L$ increases with increasing T, an expected behavior for a metal such as FePt. In agreement with Fig. \ref{fig:1}, $\rho_{L,sat}^{\mathrm{H}//\mathrm{i}}$ is expected to be higher than $\rho_{L,sat}^{\mathrm{H}\perp \mathrm{i}}$ due to the anisotropic magnetoresistance ($\rho_{L,sat}^{\mathrm{H}//\mathrm{i}} - \rho_{L,sat}^{\mathrm{H} \perp \mathrm{i}}$ = $\Delta\rho_0$ + $\Delta\rho > 0$) 
(see sketches in Fig. \ref{fig:2}(a)). Interestingly, the difference between the two orientations $\rho_L^{\mathrm{H}//\mathrm{i}} - \rho_L^{\mathrm{H} \perp \mathrm{i}}$ decreases with increasing T, both for the saturation and remanence conditions, Figs. \ref{fig:2}(b) and (c), respectively. In particular, Fig. \ref{fig:2}(b) includes the change of the magnetization with increasing T. These results indicate that the metallic-like behavior dominates the T dependence (Fig. \ref{fig:2}(a)), but an opposite and subtler behavior appears upon subtracting the two extreme orientations (H$//$i and H$\perp$i). As previously quantified, $\rho_{\perp}$ is orders of magnitude higher than $\Delta\rho_0$ + $\Delta\rho$ and rules the overall behavior of $\rho_L$ with T. Canceling its contribution, $\rho_L^{\mathrm{H}//\mathrm{i}} - \rho_L^{\mathrm{H} \perp \mathrm{i}}$ gets reduced with increasing T, following the same tendency as magnetization does (Fig. \ref{fig:2}(b)), as expected for a FM material \cite{jan_galvamomagnetic_1957-1}. 

$\rho_{L,rem}^{\mathrm{H}//\mathrm{i}}$ and $\rho_{L,rem}^{\mathrm{H} \perp \mathrm{i}}$ are measured as a function of T while applying a magnetic field of 30 Oe, at which the magnetic state does not differ from that at remanence (i.e., at zero aplied field). 
The two of them describe curves similar to the saturation ones (see Fig. S3). More precisely, their values are contained within the curves described by $\rho_{L,sat}^{\mathrm{H}//\mathrm{i}}$ and $\rho_{L,sat}^{\mathrm{H} \perp \mathrm{i}}$ as a function of T (Fig. \ref{fig:2}(a)). Interestingly, the difference between the remanence values $\rho_{L,rem}^{\mathrm{H}//\mathrm{i}}-\rho_{L,rem}^{\mathrm{H}\perp \mathrm{i}}$ as a function of T (see Fig. \ref{fig:2}(c)) changes sign. This behavior is the result of a competition between terms that contribute differently to $\rho_L$: either $\rho_{L,rem}^{\mathrm{H}//\mathrm{i}}$ increases at low T or $\rho_{L,rem}^{\mathrm{H}\perp \mathrm{i}}$ decreases. This will be discussed afterwards. 

\begin{figure}[ht!]
    \centering
    \includegraphics[width=0.5\linewidth]{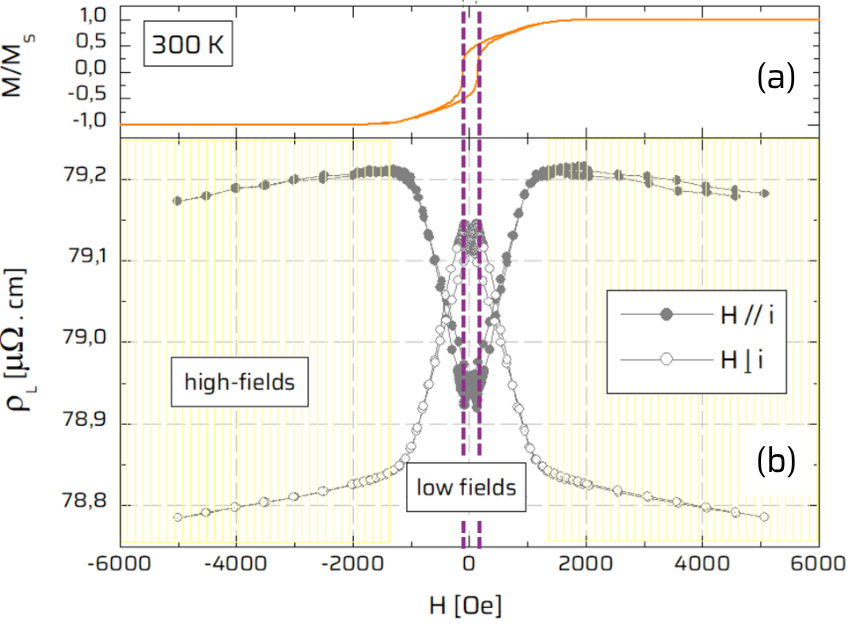}
    \caption{\textbf{Magnetization and magnetotransport properties atRT.} (a) Normalized magnetization ($\mathrm{\frac{M}{M_S}}$) and (b) longitudinal resistivity ($\rho_L$) for H$//$i and H$\perp$i, as a function of the intensity of the magnetic field (H). The two dotted lines indicate the coincidence between the coercive fields and the resistivity extrema.}
    \label{fig:3}
\end{figure}

\noindent To gain better insight into the simultaneous effects of temperature and magnetic field, a third type of experiment was performed: $\rho_L$ is measured as a function of magnetic field intensity (H) at different temperatures. Figure \ref{fig:3} shows: (a) the normalized magnetization ($\mathrm{\frac{M}{M_S}}$) and (b) the longitudinal resistivity curves ($\rho_L^{\mathrm{H}//\mathrm{i}}$ and $\rho_L^{\mathrm{H}\perp \mathrm{i}}$) as a function of H. Two regimes (high- and low-fields) and three different conditions (saturation, coercivity, and remanence) can be identified. Eqs. \ref{eq:rho-x_H-par-i_var} and \ref{eq:rho-x_H-perp-i_var} comprise the mathematical formulation derived from Eq. \ref{eq:Ohm's-law}. %

\begin{equation}
     \rho_L^{H//i}  = \left\{ \begin{array}{lc} \rho_{\perp} + \Delta\rho_0 + \Delta\rho & \mathrm{H}_{sat} < \lvert \mathrm{\vec{H}} \rvert \\ \\ \rho_{\perp} + \Delta\rho_0 + \Delta\rho \cdot f_1 & f_1 < 1 ~ , ~ 0~\mathrm{Oe} < \lvert \mathrm{\vec{H}} \rvert < \mathrm{H}_{sat} \\ \\
     \rho_{\perp} + \Delta\rho \cdot f_2 & f_2 < 1 ~ , ~ \lvert \mathrm{\vec{H}} \rvert = 0~\mathrm{Oe} 
     \end{array} \right.
    \label{eq:rho-x_H-par-i_var}
\end{equation}

\begin{equation}
     \rho_L^{\mathrm{H} \perp \mathrm{i}}  = \left\{ \begin{array}{lc} \rho_{\perp} & \mathrm{H}_{sat} < \lvert \mathrm{\vec{H}} \rvert \\ \\ \rho_{\perp} + \Delta\rho \cdot f_3 & f_3 << 1 ~ \& ~ \lvert \mathrm{\vec{H}} \rvert < \mathrm{H}_{sat} \end{array} \right.
    \label{eq:rho-x_H-perp-i_var}
\end{equation}

\begin{figure}[ht!]
    \centering
    \includegraphics[width=0.5\linewidth]{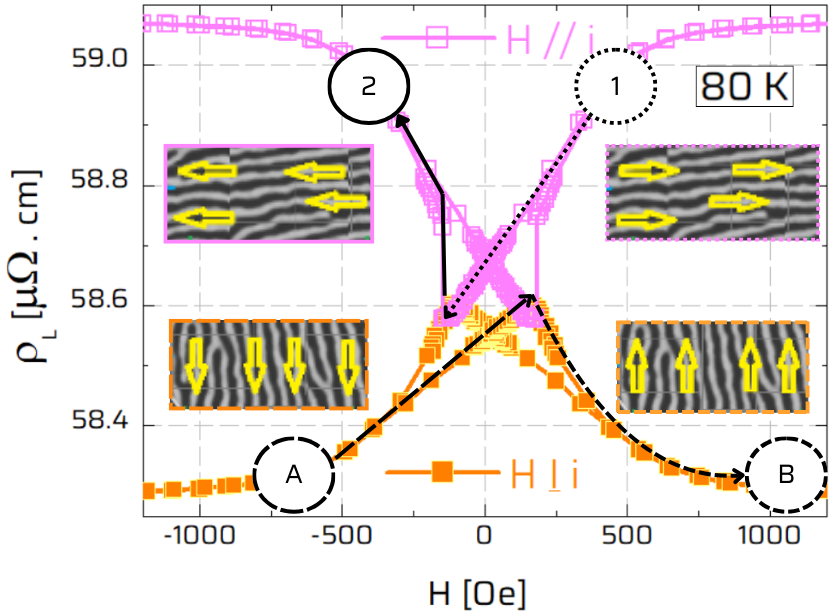}
    \caption{\textbf{Low-field longitudinal resistivity ($\rho_L$) at 80$~$K}. $\rho_L$ as a function of the field intensity (H) for H$//$i and H$\perp$i. \textbf{1} and \textbf{2} illustrate the progression of the resistivity coming from saturation towards reversal for H$//$i. \textbf{A} and \textbf{B} indicate the opposite evolution for H$\perp$i. Each pair of associated sketches represents the stripes and the relative orientation between $\rho_L$ (horizontal) and the in-plane components of $\vec{M}$ for each condition.}
    \label{fig:4}
\end{figure}

\vspace{0.2cm}
\noindent Eqs. \ref{eq:rho-x_H-par-i_var} and \ref{eq:rho-x_H-perp-i_var} predict a stable value at high fields and a variation at low fields, close to remanence, which could either be a valley or a peak, for $\rho_L^{\mathrm{H//i}}$ and $\rho_L^{\mathrm{H} \perp \mathrm{i}}$, respectively. However, additional contributions are experimentally obtained both at saturation and close to coercivity. First, the negative slope at high fields can be attributed to the electron-magnon scattering \cite{raquet_electron-magnon_2002}. This is a well-known effect reported for FM metals \cite{jan_galvamomagnetic_1957-1}. The negative MR competes with the positive MR expected for a metal, which is present but overshadowed by the dominating contribution. 
%
%
According to eq. \ref{eq:rho-x_H-perp-i_var}, $\rho_\perp$ is responsible for the high-field variation. Indeed, upon subtracting $\rho_L^{\mathrm{H}//\mathrm{i}}$ and $\rho_L^{H\perp i}$ the effect vanishes (see Fig. S4), as $\rho_{\perp}$ gets canceled in the theoretical expressions in eqs. \ref{eq:rho-x_H-par-i_var} and \ref{eq:rho-x_H-perp-i_var}. 
An aspect not properly reflected in the equations is the hysteresis observed close to the coercive fields (H$_{\mathrm{coer}}$), which is due to the presence of domains and DWs \cite{viret_spin_1996,gregg_giant_1996,levy_resistivity_1997}. 
The magnetic texture of a striped FM is strongly inhomogeneous, and that is evidenced macroscopically as the local extrema observed at H$_{\mathrm{coer}}$ (see the dotted lines in Fig. \ref{fig:3}). 
A similar measurement conducted at a low T (see Fig. \ref{fig:4}) suggests that the magnitude and abruptness of those extrema depend on the T. This will be discussed in the next section.  

\section*{Discussion}

Considering that the film possesses a micromagnetic texture which modulates the lateral distribution of current flow (Fig. \ref{fig:cAFM-MFM}(c) and (d)), we attempt to macroscopically identify the evolution of the magnetic texture by analyzing the measured magnetotransport properties (Figs. \ref{fig:1} to \ref{fig:4}). So far, regular magnetotransport measurements have been conducted, enabling the estimation of multiple parameters of the FePt film used here. Moreover, the anisotropic magnetoresistance, the resistivity increase as a function of T, and the qualitative description of the dependency as a function of the field intensity (H), have been successfully explained in terms of well-known phenomenology \cite{jan_galvamomagnetic_1957-1}. The behavior close to the coercive condition is less clear. Although there have been previous demonstrations of the macroscopic response of the resistivity in FM in these conditions \cite{gregg_giant_1996,granada_magnetotransport_2016,pianciola_magnetoresistance_2020}, none of them has succeeded in explaining the experimental results based on the lateral modulation of the conductivity. Alternatively, what is found in the literature are theoretical calculations of DWs' contributions to the in-plane resistance, estimating the ratio between the magnetoresistance associated with DWs perpendicular and parallel to the current flow, due to their ability to mix spin channels \cite{levy_resistivity_1997}.   

Careful observation of $\rho_L^{\mathrm{H}//\mathrm{i}}$ and $\rho_L^{\mathrm{H} \perp \mathrm{i}}$ as a function of H reveals a curious behavior. In Fig. \ref{fig:4}, close to coercivity, the appearance of local extrema, which are minima for $\rho_L^{\mathrm{H}//\mathrm{i}}$ and maxima for $\rho_L^{\mathrm{H} \perp \mathrm{i}}$, is observed. Labels '1' and '2' depict the evolution from a positive-field saturation to a negative one in the H$//$i condition. Labels 'A' and 'B' identify the progression from a negative to a positive H$\perp$i field. Comparing the two cases, not only does the magnitude of such extrema differ between the H$//$i and H$\perp$i measurements, but the abruptness in the resistance change (smooth in the H$\perp$i condition and steep in the H$//$i case). Moreover, the H value at which the change is observed coincides with the condition for the magnetization reversal. For the H$//$i case, this corresponds to switching the in-plane magnetization component along the current injection direction. On the contrary, when H$\perp$i, the magnetization reversal occurs transversely. We suggest that the magnitude of the change, with an accelerated variation rate in the H$//$i case, illustrates the contribution of the DWs to the longitudinal resistivity. In fact, by following the dependence of the resistivity as a function of the external field, something interesting can be observed. The correspondence of the $\rho_L^{\mathrm{H}//\mathrm{i}}$ minima with the coercive field is evident. While approaching that condition, disorder is considered to increase as a result of domains' distortion and DWs shortening and/or vanishing. Once magnetization reversal is consolidated, domains reorganize, and DWs get restored. This could be associated with the condition at which the resistivity increases again by a discrete step, allowing us to conclude that the restoration of DWs contributes to the resistivity by a sharp increase. 

Measurements as those presented in Fig. \ref{fig:3} and \ref{fig:4} were recorded between 300 K and 80 K, with increments of 20 K (see Fig. S5(a)). Overall, an evolution of the absolute values is observed, with $\rho_L$ increasing with T. The trend of the saturation resistivity values as a function of T (Fig. S5(b)) is in agreement with the results described in Figs. \ref{fig:2}(a). Looking closer to the high-fields regime, a change in the negative MR is observed: the slope gets reduced upon decreasing T (Fig. S5(c)), in agreement with the 
electron-magnon interaction \cite{jan_galvamomagnetic_1957-1}. 

\begin{figure}[ht!]
    \centering
    \includegraphics[width=\linewidth]{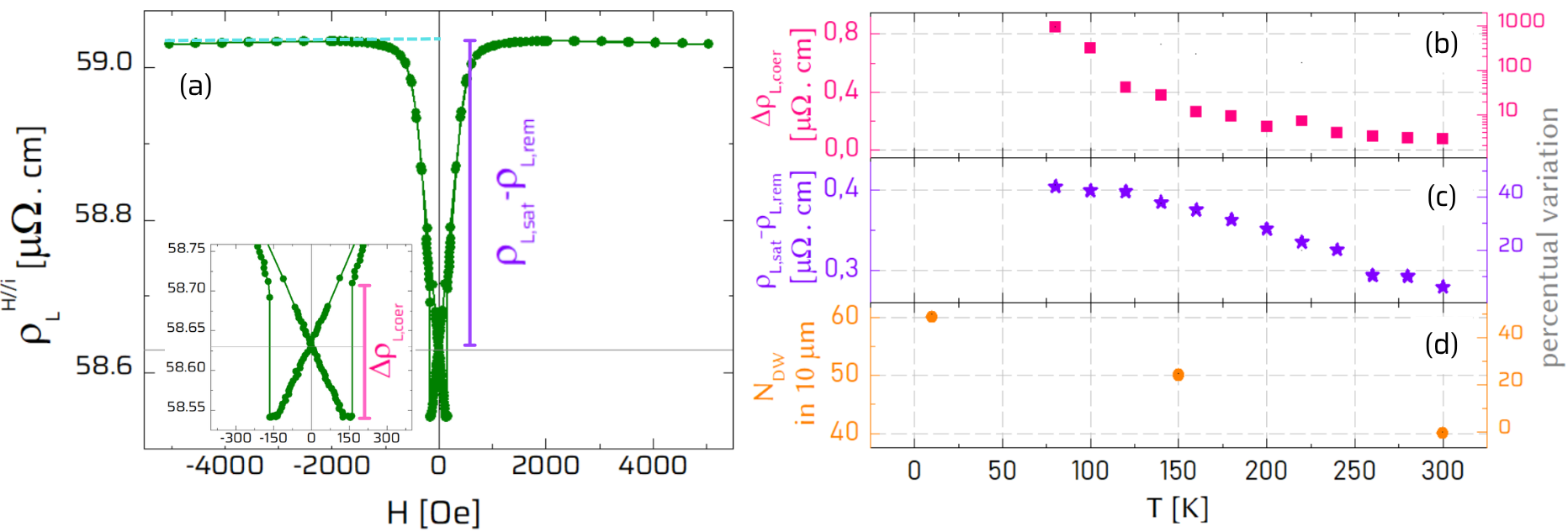}
    \caption{\textbf{Macroscopic quantities defined from magnetotransport measurements at multiple T}. (a) Longitudinal resistivity ($\rho_L$) at low-T as a function of H. $\rho_{L,sat}-\rho_{L,rem}$ is defined. The inset shows a zoomed-in area in the vicinity of H = $0~$Oe where $\Delta\rho_{L,coer}$ is defined. (b) $\Delta\rho_{L,coer}$, (c) $\rho_{L,sat}-\rho_{L,rem}$, and (d) $\mathrm{N_{DW}}$, as a function of T. In (b), (c), and (d), the left axes indicate the absolute values while the right axes indicate the percentual variation relative to the RT value.}
    \label{fig:5}
\end{figure}

To quantify the behavior of the resistivity close to the coercive field two quantities are introduced (see Fig. \ref{fig:5}(a)): the magnitude of the change in the vicinity of this condition ($\Delta\rho_{L,coer}$), and the difference between the saturation and remanent values ($\rho_{L,sat}-\rho_{L,rem}$). Fig. \ref{fig:5}(b) and (c) show their evolution as a function of T, with the low-T values being considerably higher than those recorded at RT (left axes include the net values, right axes indicate the percentual variation of the same quantities). %
%
This trend could be attributed to: \textbf{i.} a net increase of the $\vec{M}$ vector, \textbf{ii.} an increase of the longitudinal $\vec{M}$ component
, \textbf{iii.} an increase of the net amount of DWs, and/or \textbf{iv.} an increase in the contribution of each domain or DW resistivity. These explanations do not need to be mutually exclusive, and multiple mechanisms may be simultaneously at play. The following analysis examines each of them.      

\noindent Revisiting Fig. \ref{fig:2}(b), it is possible to recall that $\vert\vec{M}\vert$ (= M) does increase upon T reduction. An almost linear dependency with a negative slope of 0.46 $\frac{\mathrm{emu}}{\mathrm{cm}^3\cdot K}$ and nearly a 10 $\%$ variation is measured in the T range of interest. Additionally, our previous work on the same FePt system \cite{quinteros_thermal_2024} showed an $\mathrm{\frac{M_{rem}}{M_{sat}}}$ increase while reducing T (from 0.5 at RT to 0.6 at 10 K). This is also reflected indirectly in another macroscopically measurable quantity. Defined as the difference between the projection of the saturation at remanence and its actual value at that specific condition, $\rho_{L,sat}-\rho_{L,rem}$ (see Fig. \ref{fig:5}(a)) comprises a useful quantity to measure the ratio between the resistivity in remanence and in saturation. The increase of $\rho_{L,sat}-\rho_{L,rem}$ as a function of T reduction (see Fig. \ref{fig:5}(c)) is in agreement with the results of magnetization as a function of T \cite{quinteros_thermal_2024}. The in-plane $\vec{M}$ component increases 20 $\%$ upon T reduction while $\rho_{L,sat}-\rho_{L,coer}$ increases 60 $\%$ in the same range. Together with the concomitant increase of the net value of $\vec{M}$, this could explain a nearly 43-76 $\%$\footnote{Depending on whether the variation of the in-plane $\vec{M}$ or $\rho_{L,sat}-\rho_{L,rem}$ is combined with a net increase of M ($\vert\vec{M}\vert$).} increase. 
However, neither of these two variations justify the sign change of $\rho_{L,rem}^{\mathrm{H}//\mathrm{i}}-\rho_{L,rem}^{\mathrm{H}\perp \mathrm{i}}$ (Fig. \ref{fig:2}(c)) nor the magnitude of the $\Delta\rho_{L,coer}$ change as a function of T, from $\sim$ 0.08 at RT to $\sim$ 0.8 at 80 K (Fig. \ref{fig:5}(b)). A complementary mechanism has to be involved to explain the observed phenomenology.

The sign change of $\rho_{L,rem}^{\mathrm{H}//\mathrm{i}}-\rho_{L,rem}^{\mathrm{H}\perp \mathrm{i}}$ (Fig. \ref{fig:2}(c)) could be thought in terms of competing sources of resistivity anisotropy as it was previously observed by Ruediger \textit{et al.} \cite{ruediger_negative_1998}. The ordinary and anisotropic magnetoresistance effects possess opposite trends as a function of T, giving rise to a compensation temperature where the two effects cancel each other. However, the expected sign change is the opposite of the results presented here (Fig. \ref{fig:2}(c)), with the transverse resistivity higher than the longitudinal at low-T and the contrary at RT \cite{ruediger_negative_1998}. The additional source of anisotropy, which could reconcile our results with previous results \cite{ruediger_negative_1998}, could be attributed to the magnetic texture, i.e. domains and DWs. 

In a previous work \cite{quinteros_thermal_2024}, the change in the domain periodicity was shown using low-T magnetic force microscopy (MFM). Carefully analyzing the representativeness of the magnetic texture by considering different zones of the same film, as well as statistically processing images of multiple sizes, such periodicity was established at different T. Estimating the number of DWs ($\mathrm{N_{DW}}$) in a fixed width as the inverse of that periodicity, the values obtained for a 75 nm-thick film are displayed in Fig. \ref{fig:5}(d). Effectively, the density of DWs increases upon T reduction by nearly 50 $\%$ between RT and low T. 
Qualitatively, one might argue that the monotony of  $\Delta\rho_{L,coer}$ and $\rho_{L,sat}-\rho_{L,rem}$ with respect to $\mathrm{N_{DW}}$ illustrates the capability of the macroscopic magnetotransport measurements to identify changes in the ferromagnetic texture. Specifically, the reduction of the domains' spatial periodicity and the increase of their associated boundaries (DWs) produced by lowering T becomes evident from the behavior of a macroscopic measurement, $\rho_L^{\mathrm{H//i}}$, and in particular the newly introduced quantity referred to as $\Delta\rho_{L,coer}$ (Fig. \ref{fig:5}(d)). The magnitude of the observed change is not only measurable but even higher than the anisotropic term as obtained from Fig. \ref{fig:1}. Quantitatively, 
the variation of $\Delta\rho_{L,coer}$ as a function of T follows the same tendency as $\vec{M}$ and $\mathrm{N_{DW}}$ (increasing while T decreases). Still, it does not suffice to justify its magnitude (unless a highly non-linear dependence is proposed). This indicates that, in addition to the increase in $\mathrm{N_{DW}}$ (Fig. \ref{fig:5}(d)), the specific contribution to resistivity of each DW ($\rho_{i,DW}$) also increases while reducing temperature. Being this additional contribution orders of magnitude higher than $\mathrm{N_{DW}}$ (previously measured using magneto-force microscopy \cite{quinteros_thermal_2024}), the variation of $\rho_{i,DW}$ does not enable us to determine $\mathrm{N_{DW}}$ unambiguously from the MR measurements. Noticeably, the estimated change in $\rho_{i,DW}$ as a function of temperature appears to be considerably higher than predicted by the available models.

An unconsidered reason for the marked increase in $\Delta\rho_{L,coer}$ relates to the increase of each domain and/or DW resistivity on its own upon T reduction. As can be seen in the net increase of $\Delta\rho_{L,coer}$ and its relative variation compared to all the plausible explanations proposed (see Fig. \ref{fig:5}(b)-(d)), a significant increase of the individual contribution of the domains themselves and/or the DWs to the collective film resistivity would be the additional factor required. Preliminarily, we obtained two distinct I-V dependencies while recording cAFM (not shown), indicating high- and low-conductance regions. Although neither of them seemed to admit a linear fit, which, in turn, could be attributed to the contribution of the tip-sample junction serially added to the recording current, this does not rule out that the spatial modulation of the charge carriers impacts the overall resistivity. Although a semiconducting nature of the paths for the current flow, which would explain the resistivity increase at low temperatures, would be hard to rationalize within a purely metallic alloy, the internal rearrangement of the carriers within the film due to the variation of the quality factor is not understood yet. Whether the competition among anisotropy terms (responsible for modifying the stripes' periodicity) successfully explains the observed change remains an open question. A dedicated cAFM study would be required to support this hypothesis. Unfortunately, microscopic measurements controlling the environmental humidity content to ensure the good quality of the tip-sample contact are out of the scope of this study. Moreover, a theoretical formulation based on the specific structure of the magnetic texture in FePt films would be required to rationalize such a behavior. 



In summary, from our experimental data, it is thus possible to assure that the contribution of the magnetic microtexture to the macroscopic magnetotransport properties is sizeable and its magnitude, quantified as a newly introduced term referred to as $\Delta\rho_{L,coer}$, may even surpass the anisotropic term at low temperatures. Whether the pronounced increase of such a magnitude arises from the semiconducting nature of the individual domain walls, a modification of the quality factor (via the T-dependent balance among the anisotropy terms). or other unconsidered aspects poses a question open to further studies.

\section*{Conclusions}




Magnetoresistance measurements were performed on a Hall bar–structured FePt film with ferromagnetic stripes. Generalized Ohm's law was used to deduce the expected behavior of longitudinal resistivity in multiple configurations and conditions: a magnetic field, with an intensity varying from saturation to remanence, was applied along and across the injected current, and recorded at different temperatures. Beyond the terms considered in the mathematical formulation, contributions coming from the magnetic texture were identified and characterized.   

Through a systematic set of magnetotransport measurements, several features were directly linked to its underlying micromagnetic texture. In particular, the behavior of the longitudinal resistivity during magnetization reversal, the sign change observed between remanent states under parallel and perpendicular field configurations, and the magnitude of the resistivity discontinuity near the coercive field all reveal a strong influence of magnetic domains and domain walls. Furthermore, the analysis of the resistivity variations suggests that domain walls contribute distinctly to the overall transport properties, appearing to be more resistive than the domains themselves. Moreover, by introducing two quantities: the magnitude of the change in the vicinity of the coercive field ($\Delta\rho_{L,coer}$), and the difference between the saturation and remanent values ($\rho_{L,sat}-\rho_{L,rem}$), the impact of magnetic texture on longitudinal resistivity is represented as a function of temperature. Upon lowering T, the relative importance of the resistivity terms varies with $\Delta\rho_{L,coer}$ surpassing the RT estimation of $\rho_{\perp}$ and $\Delta\rho+\Delta\rho_0$. 

Overall, the results demonstrate that spatial modulation of the magnetic texture significantly affects macroscopic electrical properties, producing effects that are not only measurable but also substantial. Although further work is needed to quantitatively relate domain and domain wall densities to transport behavior, this study clearly establishes that microscopic magnetic inhomogeneities play a key role in determining macroscopic magnetotransport responses, highlighting their potential relevance for technological applications.

\ack{The authors kindly acknowledge Prof. Alejandro Butera for his insightful comments and the fruitful discussions. The authors are also grateful to the staff of the INN-CAB cleanroom facility for technical assistance. Technical support from Rubén E. Benavides, César Pérez, and Matías Guillén is greatly acknowledged.}

\funding{This work was partly supported by CONICET PIP 2023-2025 11220220100508CO, CONICET PIET-R 2025 29820250100057CO, PIBAA 2022–2023 28720210100099CO, the Argentine MinCyT Projects PICT-2021-I-INVI-00113 (project DISCO), and by the University of Cuyo by Grants No. 06/C035-T1, 06/80020240100271UN and 80020240100457UN.}

\roles{

\noindent \textbf{Cynthia P. Quinteros}:
Conceptualization (Lead)
Data curation (Equal)
Formal analysis (Equal)
Funding acquisition (Equal)
Investigation (Equal)
Methodology (Equal)
Project administration (Lead)
Writing - original draft (Lead)
Writing - review and editing (Equal)

\noindent \textbf{Luis Avilés-Félix}:
Conceptualization (Equal)
Data curation (Equal)
Formal analysis (Equal)
Investigation (Equal)
Methodology (Lead)
Software (Supporting)
Supervision (Equal)
Validation (Equal)
Writing - original draft (Equal)
Writing - review and editing (Equal)

\noindent \textbf{Dafne Goijman}:
Formal analysis (Equal)
Investigation (Equal)
Methodology (Equal)
Resources (Equal)
Writing - original draft (Equal)

\noindent \textbf{Lautaro Saba}:
Investigation (Equal)
Methodology (Equal)
Software (Equal)

\noindent \textbf{Diego Pérez-Morelo}:
Investigation (Equal)
Methodology (Equal)
Supervision (Equal)

\noindent \textbf{Leticia Granja}:
Conceptualization (Supporting)
Data curation (Equal)
Formal analysis (Equal)
Investigation (Equal)
Methodology (Equal)
Supervision (Equal)
Visualization (Equal)
Writing - original draft (Equal)

\noindent \textbf{Mara Granada}:
Formal analysis (Equal)
Investigation (Equal)
Methodology (Equal)
Supervision (Equal)
Validation (Equal)
Visualization (Equal)
Writing - original draft (Equal)
Writing - review and editing (Lead)

\noindent \textbf{Julián Milano}:
Conceptualization (Equal)
Formal analysis (Equal)
Investigation (Equal)
Methodology (Equal)
Supervision (Equal)
Validation (Equal)
Writing - original draft (Equal)
Writing - review and editing (Equal)

}

\data{\noindent All data relevant to supporting the findings of this study are included within the main manuscript or in the supplementary materials. For any further information, please do not hesitate to contact the authors, who will gladly provide access to additional data upon reasonable request.}

\data{\noindent The authors declare no conflict of interest.}

\printbibliography

\suppdata{

\setcounter{figure}{0}
\renewcommand{\figurename}{Fig.}
\renewcommand{\thefigure}{S\arabic{figure}}

\section*{Mathematical expressions for the resistivity terms}

\noindent General formulation of Ohm's law is as follows: 

\vspace{-0.1cm}
\begin{equation}
    \Vec{E} = \bar{\bar{\rho}} \cdot \Vec{J} \Rightarrow 
    \left[\begin{array}{c}
E_{x} \\
E_{y} \\
E_{z} \\
\end{array}\right] 
= 
\left[\begin{array}{ccc}
\rho_{xx} & \rho_{xy} & \rho_{xz} \\
\rho_{yx} & \rho_{yy} & \rho_{yz} \\
\rho_{zx} & \rho_{zy} & \rho_{zz} \\
\end{array}\right] 
\cdot 
\left[\begin{array}{c}
J_{x} \\
J_{y} \\
J_{z} \\
\end{array}\right] 
\end{equation}

\noindent However, due to the difficulty involved in obtaining an analytical expression from it, an empirical formulation is used. The simplest Ohm's law assumes that resistance (resistivity) is due to the current flow (current density) promoted by the presence of a voltage (electric field) and is therefore expressed as a linear dependence between the field and the current density, that is, $\Vec{E} = \rho_{\perp} \Vec{J}$. In contrast, the more general empirical formulation also considers contributions due to: magnetoresistive effects (ordinary and anisotropic), as well as the Hall effects (ordinary and anomalous). The above generalized empirical law is expressed in equation 1 (Main text).

To express the different components with their respective terms, it is necessary to define the respective coordinate systems. The experiments are carried out on a FePt sample delimited laterally in the shape of a Hall bar (Figure 1(a) of the Main text). This determines an ($\hat{x},\hat{y}$) plane. The sample is mounted on a sample holder, which is in turn located inside a cryostat that isolates it from its surroundings. The electromagnet used to apply the external magnetic field rotates around it. The magnet rotates in the ($\hat{x'},\hat{y'}$) plane. In this way, two reference systems are considered. The first is associated with the sample coordinates ($\hat{x},\hat{y},\hat{z}$) and the second is associated with the magnetic field ($\hat{x}',\hat{y}',\hat{z}'$). Thus, the magnetization is expressed as follows:

\vspace{-0.1cm}
\begin{equation}
    \Vec{M} = \mathrm{M} \cdot \cos(\phi_M) \cdot \sin(\theta_M) \hspace{1mm} \hat{x} + \mathrm{M} \cdot \sin(\phi_M) \cdot \sin(\theta_M) \hspace{1mm} \hat{y} + \mathrm{M} \cdot \cos(\theta_M) \hspace{1mm} \hat{z} 
    \label{eq:M}
\end{equation}

\noindent The vector formulation of the magnetization considers that the stripes may form an angle ($\phi_M$) with the $+\hat{x}$ direction (as if the initial alignment had not been completely effective) and that it may not be contained in the plane ($\theta_M$ is the angle formed by the magnetization with the ($\hat{x},\hat{y}$) plane of the sample).

The magnetic field is expressed as follows:

\vspace{-0.1cm}
\begin{equation}
    \Vec{H} = \mathrm{H} \cdot \cos(\phi_H) \cdot \sin(\theta_H) \hspace{1mm} \hat{x}' + \mathrm{H} \cdot \sin(\phi_H) \cdot \sin(\theta_H) \hspace{1mm} \hat{y}' + \mathrm{H} \cdot \cos(\theta_H) \hspace{1mm} \hat{z}'
    \label{eq:H}
\end{equation}

\noindent The vector formulation of the field assumes the possibility of varying the angle in the orientation plane to the $\hat{x}'$ axis ($\phi_H$) as well as outside of it ($\theta_H$).

The reference system associated with the Hall bar and, therefore, with the plane containing the FePt film will be used. This means that we will implement a change of basis from the system ($\hat{x}'~\hat{y}'~\hat{z}'$) to ($\hat{x}~\hat{y}~\hat{z}$).

\begin{figure}[ht!]
\centering

\begin{tikzpicture}[x = 2.0cm, y = 2.0cm, z = 2.0cm, scale = 3, tdplot_main_coords]

	\coordinate (Origin) at (0, 0, 0);
	
	%
	\draw [canvas is yx plane at z=0, black!10!white, step = 0.2] (-1, -1) grid (1, 1);
	\draw [canvas is zx plane at y=0, black!10!white, step = 0.2] (-0.6, -1) grid (1, 1);
	\draw [canvas is zy plane at x=0, black!10!white, step = 0.2] (-0.6, -1) grid (1, 1);

	%
	\draw [arrows = {}-{latex}, line width = 1.0pt, GreenDark] (Origin) -- (1, 0, 0)%
		node[anchor = east, xshift = -1mm, inner sep = \TikzInnerSep]%
			{$\VL{\hat{x}}$};
	%
	\draw [arrows = {}-{latex}, line width = 1.0pt, GreenDark] (Origin) -- (0, 1, 0)%
		node[anchor = north west, xshift = +1mm, inner sep = \TikzInnerSep]%
			{$\VL{\hat{y}}$};
	%
	\draw [arrows = {}-{latex}, line width = 1.0pt, GreenDark] (Origin) -- (0, 0, 1)%
		node[anchor = south west, xshift = +1mm, yshift = -1mm, inner sep = \TikzInnerSep]%
			{$\VL{\hat{z}}$};
    
    \centerarc[canvas is yx plane at z=0, arrows = {latex}-{}](0, 0)(400 : 500 : 0.15) 
		node[anchor = north west, xshift = -7mm, yshift = +1mm, inner sep = \TikzInnerSep]%
			{$\phi$};
    
    \centerarc[canvas is yz plane at x=-0.2, arrows = {}-{latex}](0, 0)(120 : 140 : 0.65) 
		node[anchor = south, xshift = +1mm, yshift = +3mm, inner sep = \TikzInnerSep]%
			{$\theta$};
    %
	\draw [arrows = {}-{latex}, line width = 1.0pt, red!90] (Origin) -- (0.87, 0.5, 0)%
		node[anchor = east, xshift = -1mm, inner sep = \TikzInnerSep]%
			{$\VL{\hat{x'}}$};
    %
	\draw [arrows = {}-{latex}, line width = 1.0pt, red!90] (Origin) -- (-0.49, 0.85, 0.17)%
		node[anchor = east, xshift = 5mm, inner sep = \TikzInnerSep]%
			{$\VL{\hat{y'}}$};
    %
	\draw [arrows = {}-{latex}, line width = 1.0pt, red!90] (Origin) -- (0.09, -0.15, 0.98)%
		node[anchor = east, xshift = -1mm, inner sep = \TikzInnerSep]%
			{$\VL{\hat{z'}}$};
   
    \tdseteulerzxz
    \tdplotsetrotatedcoords{\zTwoRot}{\xRot}{0}
    
   \fill[tdplot_rotated_coords, canvas is yx plane at z=0, opacity = 0.2, fill = red, draw = black] (0, 0) rectangle (1, 1);
	\fill[tdplot_rotated_coords, canvas is zx plane at y=0, opacity = 0.2, fill = red, draw = black] (0, 0) rectangle (1, 1);
	\fill[tdplot_rotated_coords, canvas is zy plane at x=0, opacity = 0.2, fill = red, draw = black] (0, 0) rectangle (1, 1);

\end{tikzpicture}
\caption{Sample's and external magnetic field's coordinate systems used to formulate eqs. \ref{eq:M} to \ref{eq:Hxyz}.}  \label{fig:sist-ref}
\end{figure}
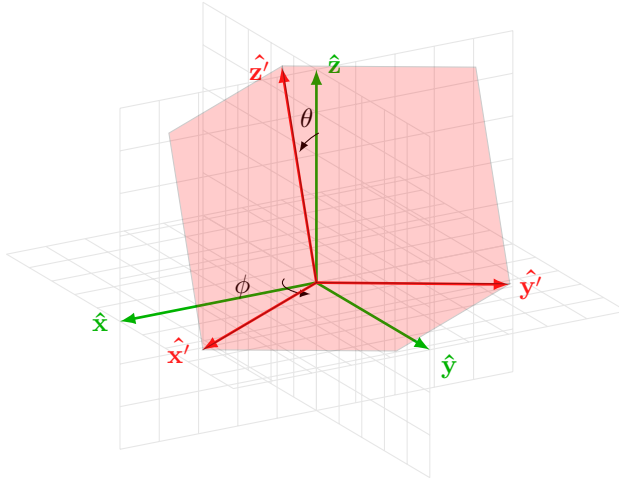

\noindent Given the degrees of freedom existing in the experimental system, a change of basis is considered between the reference system attached to the film plane ($\hat{x}~\hat{y}~\hat{z}$) and that corresponding to the rotation plane of the field ($\hat{x}'~\hat{y}'~\hat{z}'$). They differ by two rotations: one around the $\hat{z}$ axis followed by another around the $\hat{x}$ axis. This is illustrated by the following rotation matrix:

\vspace{-0.1cm}
\begin{equation}
    \left[\begin{array}{c}
\hat{x}' \\
\hat{y}' \\
\hat{z}' \\
\end{array}\right] 
= 
\left[\begin{array}{ccc}
1 & 0 & 0 \\
0 & \cos(\theta) & \sin(\theta) \\
0 & -\sin(\theta) & \cos(\theta) \\
\end{array}\right] 
\cdot
\left[\begin{array}{ccc}
\cos(\phi) & \sin(\phi) & 0 \\
-\sin(\phi) & \cos(\phi) & 0 \\
0 & 0 & 1 \\
\end{array}\right]
\cdot
\left[\begin{array}{c}
\hat{x} \\
\hat{y} \\
\hat{z} \\
\end{array}\right] 
\end{equation}

\noindent Considering the change of basis illustrated by the diagram, the following system of equations is obtained:

\vspace{-0.1cm}
\begin{equation}
\begin{array}{c}
\hat{x}' = +\cos \phi ~ \hat{x} + \sin \phi ~ \hat{y}\\
\hat{y}' = -\sin \phi \cdot \cos \theta ~ \hat{x} + \cos \phi \cdot \cos \theta ~ \hat{y} + \sin \theta ~ \hat{z} \\
\hat{z}' = +\sin \phi \cdot \sin \theta ~ \hat{x} - \cos \phi \cdot \sin \theta ~ \hat{y} + \cos \theta ~ \hat{z} \\
\end{array}
\end{equation}

\noindent To perform the replacement in the equations using a single coordinate system, the generic expression for the magnetic field is transformed to the basis associated with the sample. For simplicity, we will use the expression $\vec{H} = \mathrm{H_{x'}}~ \hat{x}'+\mathrm{H_{y'}}~ \hat{y}'+\mathrm{H_{z'}}~ \hat{z}'$ as a starting point. Replacing the inverses ($\hat{x}'~\hat{y}'~\hat{z}'$) by the change of basis, it is possible to regroup the components according to the system ($\hat{x}~\hat{y}~\hat{z}$), obtaining:

\vspace{-0.1cm}
\begin{equation}
\begin{array}{c}
\mathrm{H_x} = \mathrm{H_{x'}} \cdot \cos \phi - \mathrm{H_{y'}} \cdot \sin \phi \cdot \cos \theta + \mathrm{H_{z'}} \cdot \sin \phi \cdot \sin \theta\\
\mathrm{H_y} = \mathrm{H_{x'}} \cdot \sin \phi + \mathrm{H_{y'}} \cdot \cos \phi \cdot \cos \theta - \mathrm{H_{z'}} \cdot \cos \phi \cdot \sin \theta\\
\mathrm{H_z} = \mathrm{H_{y'}} \cdot \sin \theta + \mathrm{H_{z'}} \cdot \cos \theta \\
\end{array}
\end{equation}

\noindent which, for simplicity, is specified in one of the experimental conditions used: $\mathrm{H_{z'}} = 0$ Oe. Therefore, the above equations are rewritten as follows:

\vspace{-0.1cm}
\begin{equation}
\begin{array}{c}
\mathrm{H_x} = \mathrm{H_{x'}} \cdot \cos \phi - \mathrm{H_{y'}} \cdot \sin \phi \cdot \cos \theta \\
\mathrm{H_y} = \mathrm{H_{x'}} \cdot \sin \phi + \mathrm{H_{y'}} \cdot \cos \phi \cdot \cos \theta \\
\mathrm{H_z} = \mathrm{H_{y'}} \cdot \sin \theta \\
\end{array}
\label{eq:Hxyz}
\end{equation}

\noindent In this way, it is evident that, if there is an angle between the plane of the sample and the plane of rotation of the field, an out-of-plane effective component is observed, which can give rise to the Hall effect in the measurements.

For the sake of simplicity, instead of using the resistivity tensor, we will use the empirical formulation of Ohm's law and obtain the resistivity components in a particular direction by simply projecting the field in that direction and dividing it by the magnitude of the current density, that is, $\rho = \frac{\hat{n} \cdot \Vec{E}}{\mid\Vec{J}\mid}$. The current, in turn, is expressed as $\Vec{J} = J \hspace{1mm} \hat{x}$.

To obtain expressions for the resistivities in the different directions, the vector expressions for the magnetization and the field in the same coordinate system, equations \ref{eq:M} and \ref{eq:Hxyz}, respectively, are used. For simplicity, their simplified expressions will be used: $\vec{M} = \mathrm{M_x}~\hat{x}+\mathrm{M_y}~\hat{y}+\mathrm{M_z}~\hat{z}$ and $\vec{H} = \mathrm{H_x}~\hat{x}+\mathrm{H_y}~\hat{y}+\mathrm{H_z}~\hat{z}$.

\section*{MR as a function of the H direction ($\phi_\mathrm{H}$) in saturation}

In saturation ($\sqrt{\mathrm{H_x^2} + \mathrm{H_y^2}} > \mathrm{H}_{sat}$), the magnetization is aligned with the external field. For this reason, $\theta_M$ (i.e., the angle formed by the planes of the film and the field rotation) would be small but not zero. On the other hand, $\phi_H = \phi_M$, which means that the magnetization direction is aligned with the field direction at all times. Remembering that the field in its own reference system ($\hat{x}'~\hat{y}'~\hat{z}'$) does not have an out-of-plane component ($\theta_H = 90^{\mathrm{o}}$), and considering 
$\cos(\theta) \sim 1$\footnote{This is equivalent to thinking $\theta << 1$. However, $\sin(\theta) \sim 0$ is NOT considered since, despite being small, the contribution can be significant.} is possible to reduce the expressions to 

    \begin{equation}
        \boxed{\rho_\mathrm{L} \equiv \rho_x = \Delta\rho_0 \cdot \cos^2(\phi_\mathrm{H}+\phi) + \Delta\rho \cdot \cos^2(\phi_\mathrm{H}) + \rho_{\perp}} 
    \label{eq:rho-x_Hz'0_Rvsfi-Hsat}
    \end{equation}

    \begin{equation}
        \boxed{
        \begin{split}
        \rho_\mathrm{T} \equiv \rho_y = 2 \cdot \Delta\rho_0 \cdot \cos(\phi) \cdot \sin(\phi) \cdot \cos^2(\phi_\mathrm{H}) + \{ \Delta\rho_0 \cdot [2 \cdot \cos^2(\phi) - 1] +\\  \Delta\rho \cdot \sin^2(\theta_M) \} \cdot \sin(\phi_\mathrm{H}) \cdot \cos(\phi_\mathrm{H}) +
        \rho_H \cdot \sin(\theta) \cdot \cos(\phi_\mathrm{H}) + [ \Delta\rho_0 \cdot \cos(\phi) \cdot \sin(\phi) + \rho_{AHE} \cdot \cos(\theta_M)]
        \end{split}}
    \label{eq:rho-y_Hz'0_Rvsfi-Hsat}
    \end{equation}

\noindent First, the resistance measured in the longitudinal ($\mathrm{\frac{V_L}{i}}$) and transverse ($\mathrm{\frac{V_L}{i}}$) directions is analyzed by varying the angle ($\phi$) of the applied magnetic field ($\lvert \vec{H} \rvert = \mathrm{H} = 
8000~\mathrm{Oe}$, saturation condition) while a direct current is injected between electrodes A and B (see Fig. 2(a) of the Main). In that case, the equations \ref{eq:rho-x_Hz'0_Rvsfi-Hsat} and \ref{eq:rho-y_Hz'0_Rvsfi-Hsat} give rise to the following expressions:

\vspace{-0.1cm}
\begin{equation}
    \mathrm{\frac{V_L}{i}} = \mathrm{R_{xx}} = \color{teal}A_{xx} \color{black}\cdot \cos^2(\phi_\mathrm{H}+\color{gray}\phi\color{black}) + \color{red}B_{xx} \color{black}\cdot \cos^2(\phi_\mathrm{H}) + \color{green}C_{xx} \color{black}
\label{eq:R-JK_sat}
\end{equation}

\vspace{-1cm}



\begin{equation}
    \mathrm{\frac{V_T}{i}} = \mathrm{R_{xy}} = \color{blue}A_{xy} \color{black}\cdot \cos^2(\phi_\mathrm{H}) + \color{violet}B_{xy} \color{black} \cdot \cos(\phi_\mathrm{H}) \cdot \sin(\phi_\mathrm{H}) + \color{orange}C_{xy} \color{black} \cdot \cos(\phi_\mathrm{H}) + \color{cyan}D_{xy} \color{black}
\label{eq:R-CD_sat}
\end{equation}

\noindent The experimental data acquired at room temperature, together with the dependencies expressed by the previous equations, are presented in Figure \ref{fig:S1}.

\begin{figure}[ht!]
    \centering
    \includegraphics[width=\linewidth]{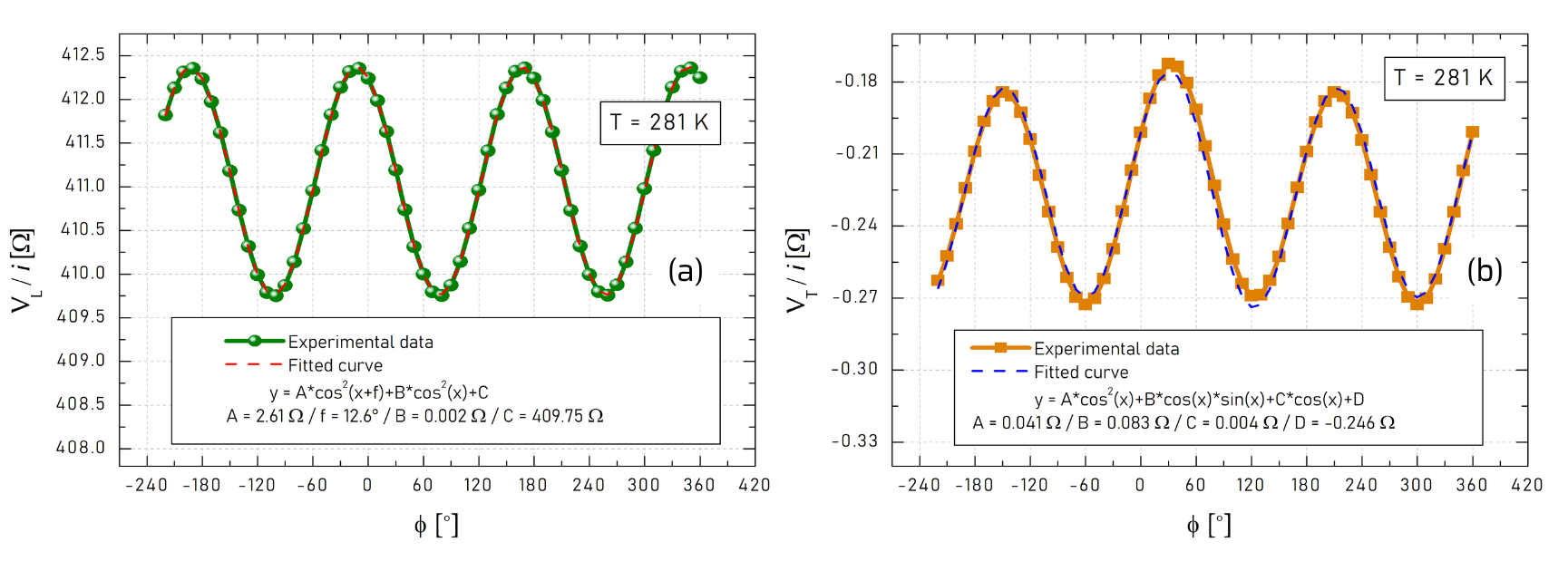}
    \caption{\textbf{Experimental magnetoresistance $@$ 281 K.} (a) Longitudinal, R$_{xx}$ and (b) transversal R$_{xy}$ magnetoresistance as a function of the direction of $\vec{H}$, $\phi_\mathrm{H}$. The expressions used to fit each dataset are expressed in eqs. \ref{eq:R-JK_sat} and \ref{eq:R-CD_sat}.}
    \label{fig:S1}
\end{figure}

\noindent Considering Eqs. \ref{eq:R-JK_sat} and \ref{eq:R-CD_sat} and the values extracted from the experimental data fits, the following system of equations is obtained. Using the minimum value and the variation of $\mathrm{\frac{V_L}{i}}$ as a function of $\cos^2(\phi_\mathrm{H})$, we have:

\vspace{-0.5cm}
\begin{equation}
    \color{teal}A_{xx} \color{black} = \Delta \rho_0 \cdot \frac{d_L}{a} = \Delta \rho_0 \cdot \frac{40 \cdot l}{a} \eqstackrel{T = 281 K}     
    2.61~\Omega 
    \label{eq:Ax}
\end{equation}

\vspace{-0.5cm}

\begin{equation}
    \color{red}B_{xx} \color{black}= \Delta \rho \cdot \sin^2(\theta_M) \cdot \frac{d_L}{a} = \Delta \rho \cdot \sin^2(\theta_M) \cdot \frac{40 \cdot l}{a} \eqstackrel{T = 281 K} 0.002~\Omega 
    \label{eq:Bx}
\end{equation}

\vspace{-0.5cm}

\begin{equation}
    \color{green}C_{xx} \color{black}= \rho_{\perp} \cdot \frac{d_L}{a} = \rho_{\perp} \cdot \frac{40 \cdot l}{a} \eqstackrel{T = 281 K} 409.75~\Omega 
    \label{eq:Cx}
\end{equation}

\vspace{-0.5cm}
\noindent where $d_L$ represents the physical length between the two electrodes used to quantify V$_\mathrm{L}$, $a$ is the cross-section area of the metallic path, and, consequently, $\frac{d_L}{a}$ comprises the factor relating R and $\rho$. Similarly, in the following equations, $d_T$ represents the electrodes' distance between which V$_\mathrm{T}$ is determined. 

\begin{equation}
    \color{gray}\phi \color{black} \eqstackrel{T = 281 K} 12.6^{\circ}
    \label{eq:fx}
\end{equation}

\vspace{-0.8cm}

\begin{equation}
    \color{blue}A_{xy} \color{black}= 2 \cdot \Delta\rho_0 \cdot \cos(\phi) \cdot \sin(\phi) \cdot \frac{d_T}{a}  = 2 \cdot \Delta\rho_0 \cdot \cos(\phi) \cdot \sin(\phi) \cdot \frac{l}{a}  \eqstackrel{T = 281 K} 0.041~\Omega
    \label{eq:Ay}
\end{equation}

\vspace{-1cm}


\begin{equation}
    \color{violet}B_{xy} \color{black}= \{ \Delta\rho_0 \cdot [2 \cdot \cos^2(\phi) - 1] + \Delta\rho \cdot \sin^2(\theta_M) \} \cdot \frac{d_T}{a} = \{ \Delta\rho_0 \cdot [2 \cdot \cos^2(\phi) - 1] + \Delta\rho \cdot \sin^2(\theta_M) \} \cdot \frac{l}{a} \eqstackrel{T = 281 K} 0.083~\Omega 
    \label{eq:By}
\end{equation} 

\vspace{-0.7cm}

\begin{equation}
    \color{orange}C_{xy} \color{black}= \rho_H \cdot \sin(\theta) \cdot \frac{d_T}{a} = \rho_H \cdot \sin(\theta) \cdot \frac{l}{a} \eqstackrel{T = 281 K} 0.004~\Omega
    \label{eq:Cy}
\end{equation}

\vspace{-1cm}


\begin{equation}
    \color{cyan}D_{xy} \color{black} = [\Delta\rho_0 \cdot \cos(\phi) \cdot \sin(\phi) + \rho_{AHE} \cdot \cos(\theta_M)] \cdot \frac{d_T}{a} = [\Delta\rho_0 \cdot \cos(\phi) \cdot \sin(\phi) + \rho_{AHE} \cdot \cos(\theta_M)] \cdot \frac{l}{a} \eqstackrel{T = 281 K} -0.246~\Omega
    \label{eq:Dy}
\end{equation}

\noindent The equations \ref{eq:Ax} to \ref{eq:Cx} allow us to obtain some physical quantities of the system. Considering that $a = l \cdot t$, that is, the cross-sectional area of the metal tracks is the track width ($l$) and the FePt thickness ($t = 75$ nm), it is possible to express:

\begin{equation}
    \left\{ \begin{array}{c} \Delta\rho_0(T) = A_{xx}(T) \cdot \frac{t}{40} = A_{xx}(T) \cdot \frac{75~\mathrm{nm}}{40} \\ \\ \Delta\rho(T) = \frac{B_{xx}(T)}{\sin^2(\theta_M)} \cdot \frac{t}{40} = \frac{B_{xx}(T)}{\sin^2(\theta_M)} \cdot \frac{75~\mathrm{nm}}{40} \\ \\ \rho_{\perp}(T) = C_{xx}(T) \cdot \frac{t}{40} = C_{xx}(T) \cdot \frac{75~\mathrm{nm}}{40} \\ \\ \Delta\rho_0(T) = \frac{A_{xy}(T)}{2 \cdot \cos(\phi)} \cdot \sin(\phi) \cdot \frac{t}{40} = \frac{A_{xy}(T)}{2 \cdot \cos(\phi) \cdot \sin(\phi)} \cdot \frac{75~\mathrm{nm}}{40} \\ \\ \rho_H(T) = \frac{C_{xy}(T)}{\sin(\theta)} \cdot \frac{t}{40} = \frac{C_{xy}(T)}{\sin(\theta)} \cdot \frac{75~\mathrm{nm}}{40} \end{array} \right.
\end{equation}

\noindent Thus, it is possible to obtain the values of the different resistivity values (as expressed in the Main text).




\newpage
\section*{$\rho_{L,rem}^{\mathrm{H}//\mathrm{i}}$ and $\rho_{L,rem}^{\mathrm{H}\perp\mathrm{i}}$ vs H as a function of T}

\begin{figure}[ht!]
    \centering
    \includegraphics[width=0.8\linewidth]{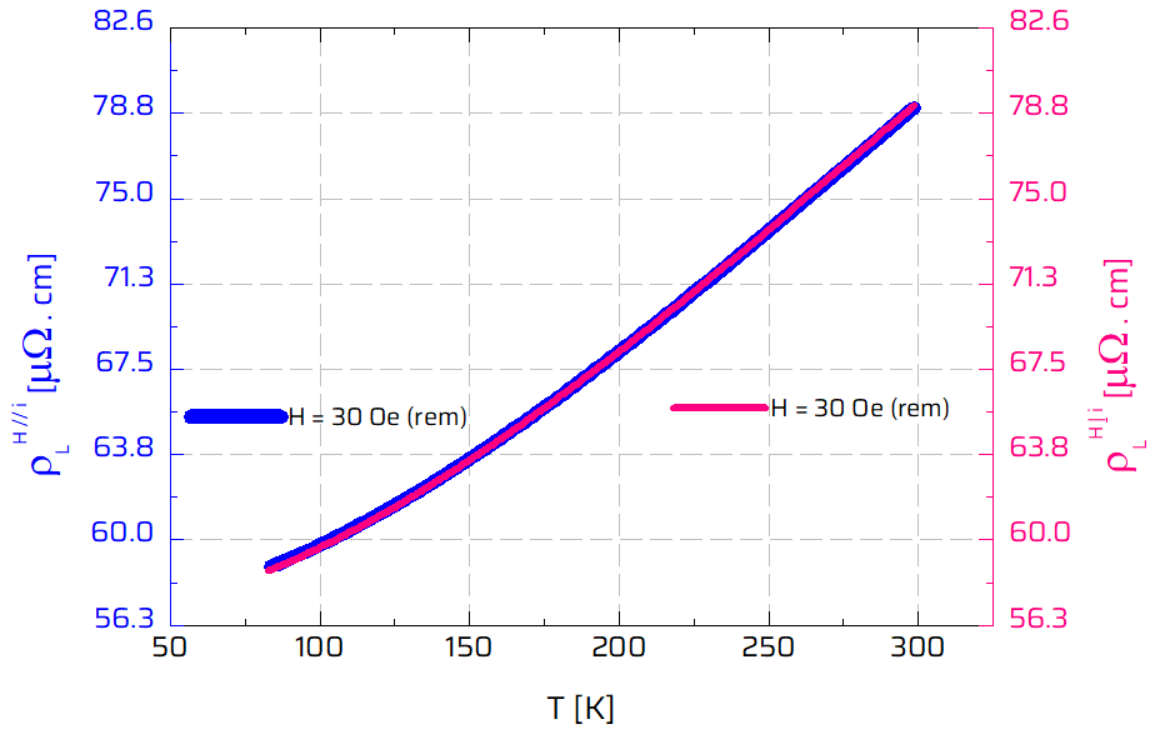}
    \caption{\textbf{Longitudinal resistivity ($\rho_L$) as a function of T.} (a) $\rho_L$ in remanence (H = 30 Oe) with H$//$i or H$\perp$i, left and right axis, respectively.}
    \label{fig:rhoLrem}
\end{figure}{}

\section*{Cancellation of the electron-magnon interaction at high fields}

\begin{figure}[ht!]
    \centering
    \includegraphics[width=0.8\linewidth]{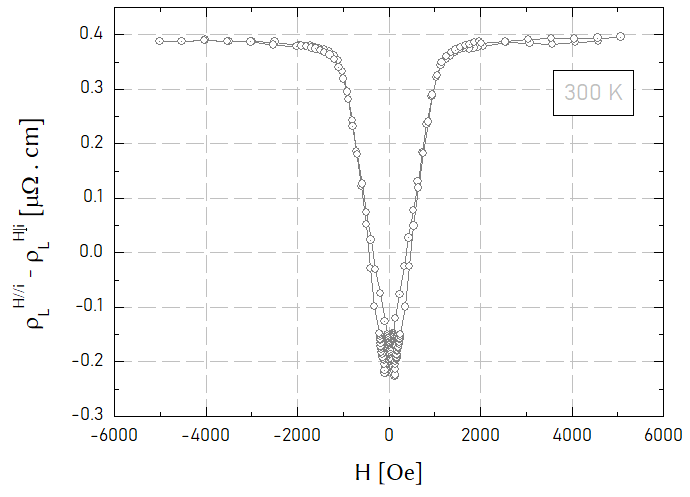}
    \caption{\textbf{$\rho^{\mathrm{H}//\mathrm{i}}_L-\rho^{\mathrm{H}\perp\mathrm{i}}_L$ as a function of H}. Subtracting the measurements applying H$//$i and H$\perp$i allows us to remove the dependence on $\rho_{\perp}$. Disappearance of the negative MR at high fields allows us to attribute it to $\rho_{\perp}$.}
    \label{fig:S2}
\end{figure}

\newpage
\section*{$\rho_L^{\mathrm{H}//\mathrm{i}}$ and $\rho_L^{\mathrm{H}\perp\mathrm{i}}$ vs H at different T}

\begin{figure}[ht!]
    \centering
    \includegraphics[width=\linewidth]{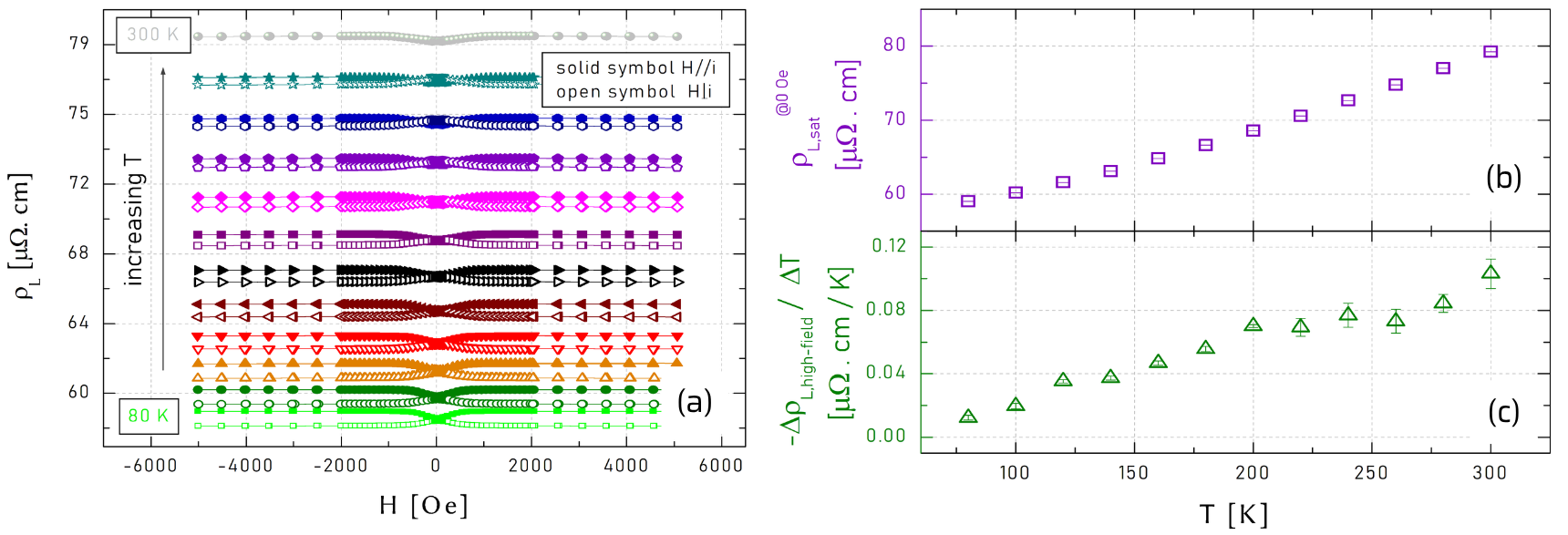}
    \caption{\textbf{$\rho_L^{\mathrm{H}//\mathrm{i}}$ and $\rho_L^{\mathrm{H}\perp\mathrm{i}}$ vs H at different T}. (a) $\rho_L$ as a function of H measured for T = $[80; 300]$ K. (b) $\rho^{@0~Oe}_{L,sat}$ and (c) $\frac{\Delta\rho_{L,high-field}}{\Delta \mathrm{T}}$ as a function of T. $\rho^{@0~Oe}_{L,sat}$ is the saturation value obtained as a projection $@$ 0 Oe (to get rid of the negative MR at high-fields). $\frac{\Delta\rho_{L,high-field}}{\Delta \mathrm{T}}$ is the high-field MR slope. As T decreases, its magnitude decreases as expected for the attenuation of the electron-magnon scattering at low T.}
    \label{fig:S3}
\end{figure}

}

\end{document}